\documentclass[prb,twocolumn,showpacs,superscriptaddress]{revtex4-1}

\usepackage{graphicx}% Include figure files
\usepackage{dcolumn}% Align table columns on decimal point
\usepackage{bm}% bold math

\begin{document}

\title{Measurement of the effect of quantum phase-slips in a Josephson junction chain}

\author{I. M. Pop}
\affiliation{Institut N\'eel, C.N.R.S. and Universit\'e Joseph Fourier, BP 166, 38042 Grenoble-cedex 9, France}
\author{I. Protopopov}
\affiliation{L. D. Landau Institute for Theoretical Physics, Kosygin str. 2, Moscow 119334, Russia}
\affiliation{Institut fuer Nanotechnologie, Karlsruher Institut fuer Technologie, 76021 Karlsruhe, Germany}
\author{F. Lecocq}
\affiliation{Institut N\'eel, C.N.R.S. and Universit\'e Joseph Fourier, BP 166, 38042 Grenoble-cedex 9, France}
\author{Z. Peng}
\affiliation{Institut N\'eel, C.N.R.S. and Universit\'e Joseph Fourier, BP 166, 38042 Grenoble-cedex 9, France}
\author{B. Pannetier}
\affiliation{Institut N\'eel, C.N.R.S. and Universit\'e Joseph Fourier, BP 166, 38042 Grenoble-cedex 9, France}
\author{O. Buisson}
\affiliation{Institut N\'eel, C.N.R.S. and Universit\'e Joseph Fourier, BP 166, 38042 Grenoble-cedex 9, France}
\author{W. Guichard}
 \affiliation{Institut N\'eel, C.N.R.S. and Universit\'e Joseph Fourier, BP 166, 38042 Grenoble-cedex 9, France}

\date{\today }
\maketitle

\textbf{The interplay between superconductivity and Coulomb interactions has been studied for more than twenty years
now\cite{Chow,Anderson,Takahide,Miyazaki,Haviland,Orr,Sharifi,Bradley,Matveev_PRL02,Glazman1997,Finkel,Fazio,Abeles}. In low-dimensional systems, superconductivity degrades in the presence of Coulomb repulsion: interactions tend to suppress fluctuations of charge, thereby increasing fluctuations of phase. This can lead to the occurrence of a superconducting-insulator transition, as has been observed in thin superconducting films\cite{Haviland,Orr}, wires\cite{Sharifi} and also in Josephson junction arrays\cite{Chow,Anderson,Takahide,Miyazaki}. The latter are very attractive systems as they enable a relatively easy control of the relevant energies involved in the competition between superconductivity and Coulomb interactions. Josephson junction chains have been successfully used to create particular electromagnetic environments for the reduction of charge fluctuations\cite{corlevi,Devoret,Glazman2009}. Recently, they have attracted interest as they could provide the basis for the realization of a new type of topologically protected qubit\cite{Gershenson,Pop} or for the implementation of a new current standard \cite{Guichard}. Here we present measurements that show clearly the effect of quantum phase slips on the ground state of a Josephson junction chain. We tune in situ the strength of quantum phase fluctuations and obtain for the first time an excellent agreement with the tight-binding model initially proposed by Matveev et al.\cite{Matveev_PRL02}. }

\begin{figure}[htbp]
\includegraphics[width=8cm]{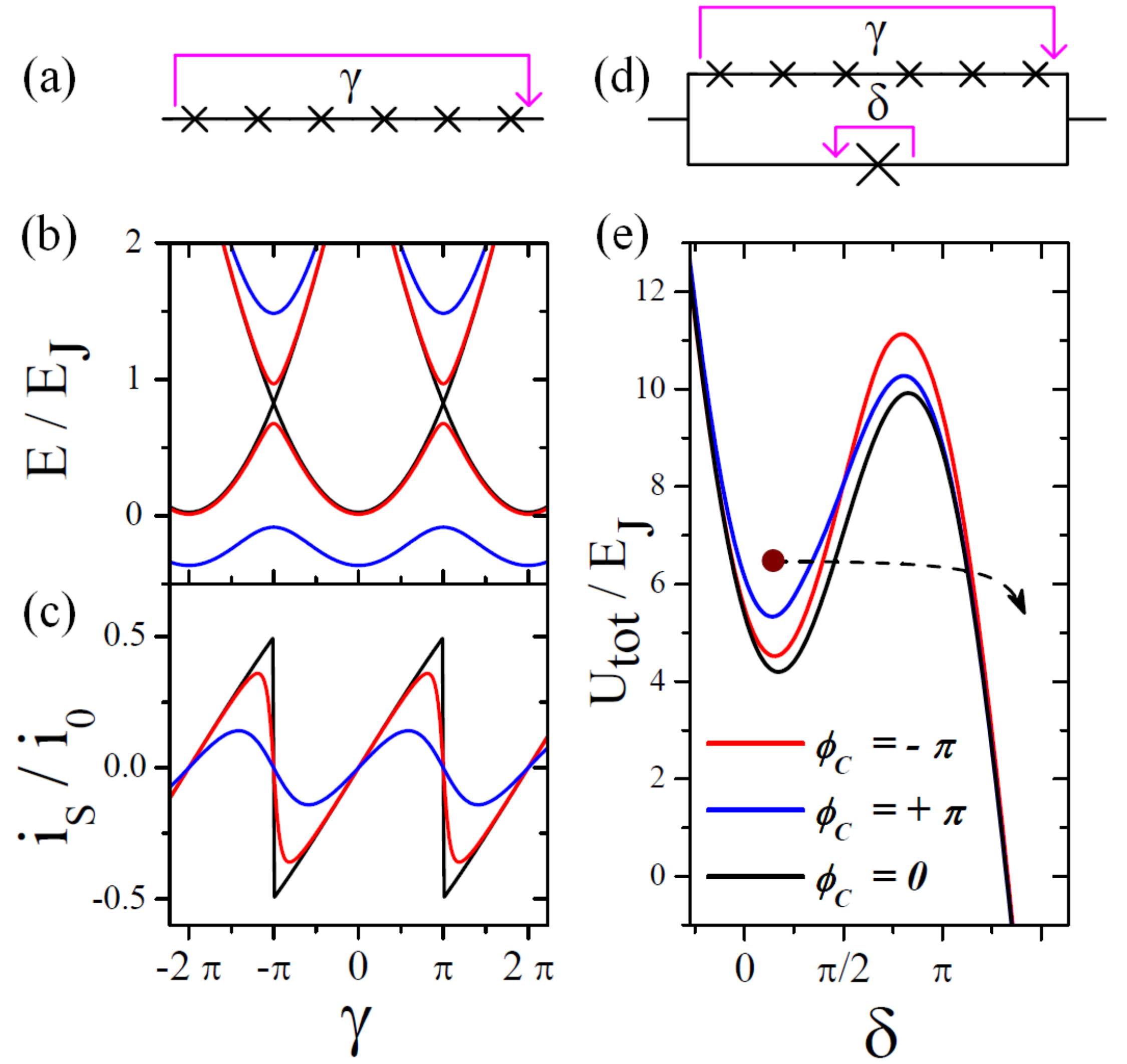}
\caption{ (a) Schematic picture of the phase-biased Josephson junction chain. (b) Energy levels of a Josephson junction chain with $N=6$ as a function of bias phase $\gamma$ for different ratios $E_{J}/E_{C}$. For $E_{J}/E_{C}=20 $ (black lines) no splitting is visible at the crossing points. For $E_{J}/E_{C}=3$ (red lines) a gap emerges that increases rapidly with decreasing $E_{J}/E_{C}$. Blue lines show the energy levels for $E_{J}/E_{C}=1.3$. For each $E_{J}/E_{C}$, the two lowest-lying states have been calculated by numerical diagonalization of the hamiltonian (\ref{hamarray}). (c) Current-phase relation for the ground state $E_{g}(\gamma)$ for the same $E_{J}/E_{C}$ ratios as in (a). The supercurrent is calculated from the derivative of the energy band: $i_{S}=\frac{2e}{\hbar} \frac{\partial E_{g}}{\partial \gamma}$. The chain current is reported in units of the critical current of a single chain junction $i_{0}=\frac{2e}{\hbar} E_{J}$. (d) Schematic picture of the chain shunted by the read-out junction. (e) Escape potential for the Josephson junction chain in parallel with the read-out junction for three different flux-biases $\phi_{C}$ in the read-out loop (see Fig. \ref{ExpSetup}). The ground-state of the chain clearly modifies the escape potential of the read-out junction.}
\label{model}
\end{figure}

The Hamiltonian for the theoretical description of superconducting circuits can be conveniently obtained by applying Devoret's circuit theory\cite{Devoretcircuit}. Here, each electrical element such as an inductance, a capacitor or the Josephson element can add a degree of freedom. In the case of circuits with a small number of electrical elements, a complete analytical description that takes into account all degrees of freedom can be obtained. However, when the  circuits contain an increasing number of elements, as for example Josephson junction chains, even numerical solutions of the problem become difficult to obtain when taking into account all degrees of freedom. Nevertheless our measurements demonstrate that the ground state of a \textit{phase-biased} Josephson junction chain (see Fig. \ref{model}(a)) can be described by a single degree of freedom. Although the chain is a multi-dimensional object, the effect of quantum phase-slips can be described by a single variable $m$, that counts the number of phase-slips in the chain.

We start by giving a short introduction on the low-energy properties of a Josephson junction
chain which have been studied in terms of quantum phase slips by Matveev et al.~\cite{Matveev_PRL02}. Let us consider the Josephson junction chain depicted in Fig. \ref{model}(a). The chain contains $N$ junctions and is biased with a phase $\gamma$.
We denote $E_J$ the Josephson energy of a single junction and $E_C=\frac{e^2}{2C}$ its charging
energy. Here we consider $E_J\gg E_C$. Let $Q_i$ be the charge on each junction and $\theta_i$ the phase difference.
In the nearest-neighbor-capacitance limit the Hamiltonian can be written as:
\begin{equation}
H = \sum _{i=1}^N [4 E_C (Q_i/2e)^2 + E_J(1 - \cos \theta_i)] \mbox{ ; }\sum_{i=1}^N \theta_i
= \gamma.
\label{QHamiltonian}
\end{equation}

Ignoring the charging energy for the moment, we find the classical ground state, that satisfies the constraint on the phase $\sum_{i=1}^N \theta_i$, by minimizing the Josephson coupling energy. The
corresponding phase configuration can be easily found for sufficiently large $N$ $(N \gg 1)$ and it is given
by $\theta_i = \gamma/N$. The resulting Josephson energy hence reads $E_0 = E_J \gamma^2 /2N$ and the chain is equivalent to a large inductance. If a phase-slip occurs on one of the junctions, say the $j$th junction, then $\theta_j \to \theta_j + 2\pi$. Since the Josephson energy is periodic in $\theta_j$, a phase-slip of $2\pi$ does not change the Josephson energy of the junction $j$. However, the constraint $\sum _i \theta_i = \gamma$ is violated after
such a phase-slip event. Therefore the phase difference $\theta_i$ over all other junctions changes a
little from $\gamma/N$ to $(\gamma-2 \pi)/N$ in order to accommodate the bias constraint.  A phase-slip on a \emph{single} junction leads to a collective response of \emph{all} junctions. Consequently the Josephson energy of the entire chain changes
from $E_0 = E_J \gamma^2 /2N$ to $E_1 = E_J (\gamma - 2\pi )^2 /2N$. One can show similarly that the classical energy needed to accommodate $m$ phase-slips without violating the constraint is given by $E_m = E_J (\gamma - 2\pi m)^2 /2N$. Therefore, the classical ground state energy of the chain, $min\left\{E_{m}\right\}$, consists of shifted parabolas that correspond respectively to a fixed number $m$ of phase-slips (see Fig. \ref{model}(b)). For the special values $\gamma = \pi (2 m +1)$ the energies $E_m$ and $E_{m+1}$ are degenerate.

Taking now into account the finite charging energy $E_C$, quantum phase-slips can lift the degeneracy at the points $\gamma = \pi (2 m +1)$. In the limit $E_J\gg E_C$, the hopping element for the quantum phase-slip can be approximated by\cite{Likharev_85,Averin}: $v = 16 \sqrt{E_J E_C/\pi} (E_J/2E_C)^{1/4} e^{-\sqrt{8 E_J/E_C}}$. Since a phase-slip can
take place on any of the $N$ junctions, the hopping term between the two states $|m\rangle$ and $|m+1\rangle$ is given by $Nv$. Therefore, using a tight-binding approximation, the total Hamiltonian for the chain is given by:
\begin{equation}
H|m\rangle = E_{m}|m\rangle - Nv\left[|m-1\rangle+|m+1\rangle\right].
\label{hamarray}
\end{equation}

Fig. \ref{model}(b) shows the numerical calculation of the two lowest eigenenergies of this Hamiltonian for three different ratios $\frac{E_{J}}{E_{C}}=20$, $3$ and $1.3$ in the case of a 6-junction chain. Fig. \ref{model}(c) shows the corresponding current-phase relation of the chain in the ground state. The chain's supercurrent is obtained by the calculation of the derivative of the ground state energy $E_g$: $i_{S}=\frac{2e}{\hbar} \frac{\partial E_{g}}{\partial \gamma}$.
For large values of $E_J/E_C$, quantum phase fluctuations are very small ($v \sim  0$) and the current-phase relation has a sawtooth-like dependence with a critical current that is approximatively $N/\pi$ times smaller than that of a single junction of the chain. We call this regime the "classical" phase-slip regime. When quantum phase fluctuations increase, i.e. $E_J/E_C$ decreases, the current-phase relation becomes sinusoidal and the critical current becomes exponentially suppressed with $N$ and $E_J/E_C$\cite{Matveev_PRL02}.
\begin{figure}[htbp]
\includegraphics[width=9cm]{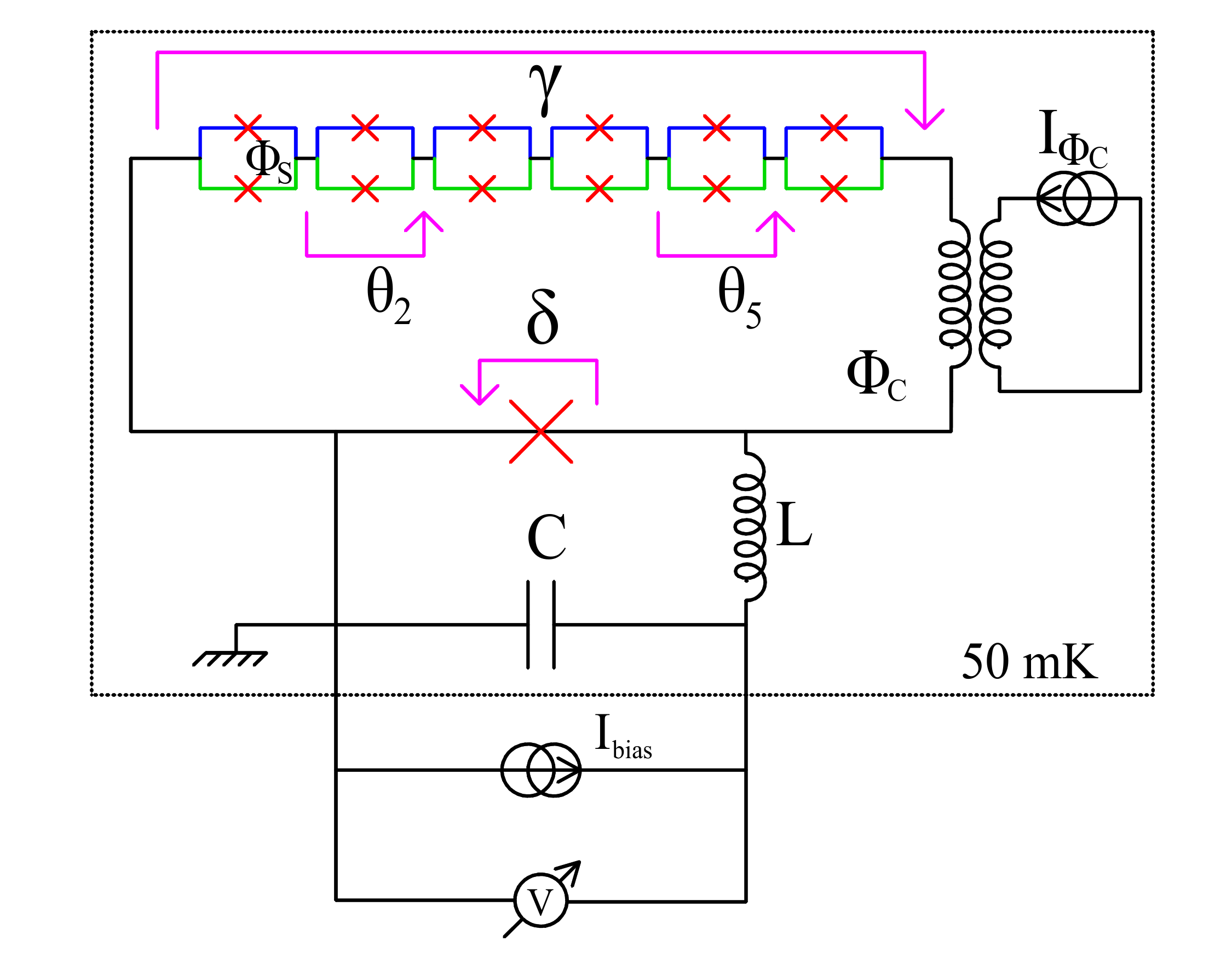}
\caption{ Measurement circuit. The 6-SQUID chain is inserted in a superconducting loop. The flux $\Phi_{C}$ created by on chip coils controls the phase difference $\gamma$ over the chain. The flux $\Phi_{S}$ through the SQUIDs can be controlled independently by a second coil. We denote the phase difference over the read-out junction $\delta$.}
\label{ExpSetup}
\end{figure}
\begin{table}[htdp]
\centering
\begin{center}
\begin{tabular}{|c|c|}
\hline
\textbf{Read-out junction} & \textbf{SQUID at $\phi_S = 0$}  \\
\hline
 $S^{RO}=(121 \pm 5) 10^3 nm^2$  &  $S^{SQ}=(30 \pm 2) 10^3 nm^2$ \\
\hline
 $C^{RO}=5.8 \pm 0.2 fF$ & $C=1.4 \pm 0.1 fF$ \\
\hline
 $R^{RO}_N=968 \pm 5 \Omega$ & $R^{SQ}_N=3800 \pm 450 \Omega$  \\
\hline
 $I^{RO}_C=330 \pm 2 nA$ & $I^{SQ}_C=83 \pm 9 nA$   \\
\hline
\end{tabular}
\end{center}
\caption{Parameters of the sample: size, capacitance, normal-state resistance and critical current of the read-out junction and a single SQUID of the chain. The critical current variance for the junctions in the chain is estimated to be smaller than $4\%$.}
\label{TableParam}
\end{table}

To measure the effect of quantum phase-slips on the ground-state of a Josephson junction chain we have studied a chain of 6 junctions. Our measurement setup and the junction parameters are presented in Fig. \ref{ExpSetup} and Table \ref{TableParam}. Each junction in the chain is realized by a SQUID in order to enable tunable Josephson coupling $E_{J}$. In this way we can tune \emph{in situ} the $E_J/E_C$ ratio by applying a uniform magnetic flux $\Phi_{S}$ through all SQUIDs, and consequently we can control the strength of quantum phase fluctuations. For our measurements we placed this chain in a closed superconducting loop, threaded by the flux $\Phi_{C}$, containing an additional shunt Josephson junction that is used for the read-out of the chain state. The flux $\Phi_{C}$ enables the control of the bias phase $\gamma=\Phi_{C}-\delta$ over the chain.

\begin{figure}[htbp]
\includegraphics[width=9cm]{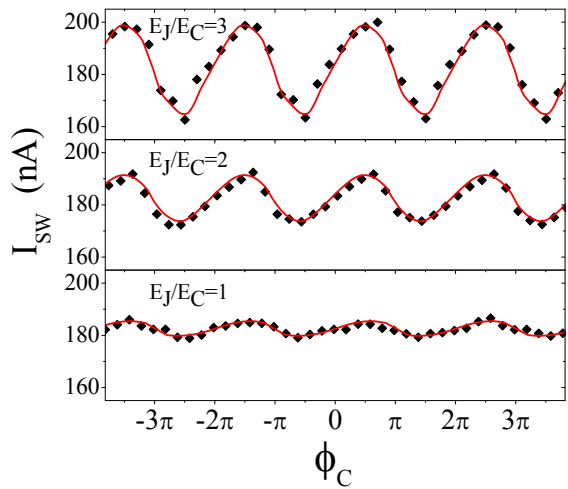}
\caption{Measured switching current (black points) as a function of $\phi_{C}$ over the chain for three different $E_{J}/E_{C}$ ratios. The measurement noise for each point is about $0.2$ nA. The red lines represent theoretical calculations for the switching current using formulas (\ref{switching}) and (\ref{TunnelCoefALL}).}
\label{CurrentPhase}
\end{figure}

We have measured the switching current of the entire Josephson junction circuit containing both the chain and the read-out junction. The switching current was determined from the switching probability at $50\%$. The escape probability as a function of bias current $I_{bias}$ shows the usual "S-form" with a width of $\approx 20$ nA. We apply typically $10000$ bias current pulses of amplitude $I_{bias}$ and duration $\Delta t \lesssim 1\mu s$, and measure the
switching probability as the ratio between the number of switching events and the total number
of pulses. The results of the switching current measurements as a function of flux $\Phi_{C}$ are shown in Fig. \ref{CurrentPhase}. From these switching current measurements we deduce the effect of quantum phase-slips on the ground state of the chain.

The measured switching current corresponds to the escape process out of the total potential energy $U_{tot}$ containing the contributions of the read-out junction and the chain:
\begin{equation}
U_{tot}(\delta,\Phi_{C})=E^{RO}_{J}\cos(\delta)+E_{g}(\Phi_{C}-\delta)-\frac{\hbar}{2e}I_{bias}\delta.
\label{Vtotal}
\end{equation}
Here $E_{g}$ is the ground state of the 6-SQUID chain calculated by solving the Hamiltonian (\ref{hamarray}). As $E^{RO}_{J} \gg E_{g}$ the main component in $U_{tot}$ is the potential of the current-biased read-out junction $E^{RO}_{J}\cos(\delta)-\frac{\hbar}{2e}I_{bias}\delta$. Fig. \ref{model}(e) shows the escape potential at constant bias-current for three different flux values $\phi_{C}$ corresponding to three different biasing phases $\gamma$ over the chain. Let us point out that the position of the minimum of the potential $U_{tot}$ is in good approximation independent of the value of the flux $\phi_{C}$. Therefore the bias phase difference $\gamma$ over the chain depends only on the flux $\phi_{C}$. As a consequence, the $\phi_{C}$ dependence of the measured switching current results from the $\gamma$ dependence of the chain's ground state.

The escape from the potential $U_{tot}$ occurs via Macroscopic Quantum Tunneling (MQT).
The MQT rate for an arbitrary potential can be calculated in the limit of weak tunneling using the dilute instanton-gas approximation\cite{DiluteInstanton}. Within this model, the escape rate $\Gamma$ out of the washboard potential $U_{tot}(\gamma)$ reads\cite{TheseNicolasDidier}:
\begin{equation}
\begin{array}{c}
\Gamma=A\exp\left[-B\right],
\\
\\
$where A and B are given by:$\\
\\
A= \sqrt{ \frac{\hbar \: {\omega_{0}}^3}{8\pi \hbar} } \sigma e^{I} $ with $
I=2 \int^{\sigma}_{0}{\sqrt{\frac{\hbar^{2} \: U_{tot}(x)}{4 \: E^{RO}_{C}}} dx}
\\
\\
B=\int^{\sigma}_{0}{\left[{ \sqrt{\frac{\hbar^2 \: {\omega_{0}}^2}{16 \: E^{RO}_{C} \: U_{tot}(x)}}-\frac{1}{x} }\right]dx}.
\end{array}
\label{TunnelCoefALL}
\end{equation}
We have denoted by $\sigma$ the width of the barrier and by $x$ the phase coordinate measured from the minimum of the washboard potential. The plasma frequency is $\omega_{0}=\sqrt{\frac{8 \: E^{RO}_{C} \: U_{tot}''(0)}{\hbar}}$, where $E^{RO}_{C}$ is the charging energy of the read-out junction.

Knowing the escape rate $\Gamma$, we can calculate the switching probability:
\begin{equation}
P(I_{bias})=1-\exp\left[-\Gamma(I_{bias}) \: \Delta t\right].
\label{switching}
\end{equation}
The results are shown as red lines in Fig. \ref{CurrentPhase}. The theory fits very well both in amplitude and shape the oscillations of the measured switching current. Let us point out that we have used the nominal values for $E_J$ and $E_C$ calculated from the characteristics of the sample indicated in Table \ref{TableParam}. The normal-state resistance for a single chain junction has been deduced from the measured normal-state resistance of the read-out junction by considering the size ratio between the two. We evaluate the precision of the determination of $E_J$ and $E_C$ to be in the range of $\pm10\%$. This error bar on $E_J$ and $E_C$ yields an uncertainty $\delta N =1$ for the effective number of junctions $N$ contributing to the phase-slip amplitude. This confirms the good homogeneity of our junctions and the collective nature of the phase-slip events.

\begin{figure}[htbp]
\includegraphics[width=9cm]{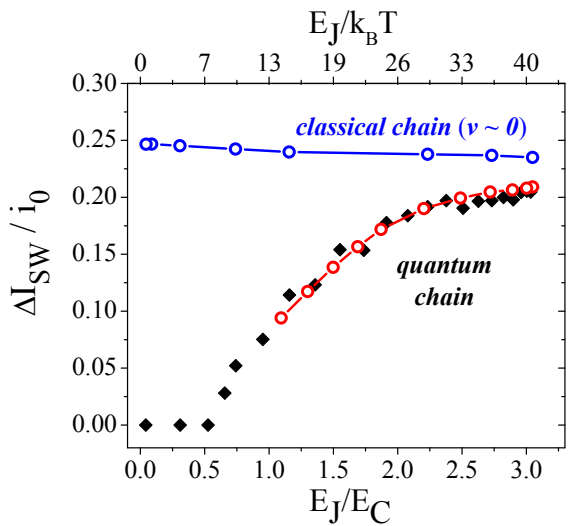}
\caption{ Comparison between the measured (black points) and the calculated (red empty circles) switching current amplitude as a function of the $E_{J}/E_{C}$ ratio. Note that the switching current amplitude is divided by the flux dependent critical current of a single SQUID $i_{0}$, in order to reveal the effect of quantum-phase fluctuations. The top curve (blue empty circles) shows the theoretical calculation of the switching current amplitude in the absence of quantum phase fluctuations. The lines are guides for the eye.}
\label{AmplDep}
\end{figure}
Fig. \ref{AmplDep} shows the measured switching current amplitude $\Delta I_{SW}$ and the corresponding theoretical calculations as a function of  $E_{J}/E_C$. For each measurement, $E_{J}$ has been calculated using the flux dependence of the SQUID's Josephson coupling: $E_{J}(\Phi_{S})=\frac{\hbar}{2e} i_{0}(\Phi_{S})$ with $i_{0}(\Phi_{S})=I^{SQ}_{C}\cos\left(\pi \frac{\Phi_{S}}{\Phi_{0}}\right) $. In order to distinguish between the suppression of the switching current that is due to quantum phase fluctuations and the one that is simply due to the well-known cancellation of the SQUID's critical current as a function of flux, we plot the switching current amplitude \emph{divided} by the critical current of a single SQUID $i_{0}$. We see that the measured switching current amplitude follows very well the predicted theoretical suppression of the switching current oscillations in the presence of quantum phase fluctuations. From our measurements we can also deduce the strength of the quantum phase-slip amplitude. With decreasing $E_J/E_C$ ratio from 3 to 1 the quantum phase-slip amplitude increases from $0.8$ GHz to $2.7$ GHz. In addition, in Fig. \ref{AmplDep} we have plotted for comparison the calculation for the switching current amplitude in the case when quantum phase fluctuations would be negligibly small: $v\sim0$. As expected, we get a practically flat dependence on $E_{J}/E_{C}$.

Further on, the upper x-axis of Fig. \ref{AmplDep} shows the ratio $E_J/k_B T$ of the Josephson energy with respect to the thermal energy at $T=50mK$. Since $E_J \gg k_B T$, thermal fluctuations are excluded to explain the suppression of the switching current with decreasing $E_J/E_C$. Additional measurements (not shown here) reveal a constant switching current amplitude up to a temperature of $T=100mK$.

In conclusion, we present for the first time a detailed experimental characterization of the effect of quantum phase-slips on the ground-state of a Josephson junction chain. These phase-slips are the result of fluctuations induced by the finite charging energy of each Josephson junction in the chain. The experimental results can be fitted in very good agreement by considering a simple tight-binding model for the phase-slips\cite{Matveev_PRL02}. Our measurements also show that a JJ chain under phase bias constraint can behave in a collective way very similar to a single macroscopic quantum object.

These results open the way for the use of quantum phase-slips in JJ networks for the implementation of a new current standard, the observation of Bloch oscillations \cite{Guichard}, the fabrication of topologically protected qubits \cite{topprot} and the design of new superconducting circuit elements.\\

\vspace{0.3cm}
\textbf{Methods}
\vspace{0.3cm}

The circuit was fabricated on a Si/SiO$_{2}$ substrate and the Al/AlO$_{x}$/Al junctions were obtained using standard shadow evaporation techniques. The aluminum oxide was obtained by natural oxidation in a controlled $O_{2}$ atmosphere. The sample was mounted in a closed copper block which was thermally connected to the cold plate of a dilution refrigerator at $50 mK$. All lines were strongly filtered by low-pass filters at the cryostat entrance and by thermocoaxial cables and $\pi$ filters at low temperatures.

The switching current $I_{SW}$ of the circuit is obtained performing the following sequence. We use a series of $M$ current steps of equal amplitude $I_{bias}$ to bias the junction. We count the number of transitions to the voltage state $M_{SW}$ and thus obtain the value of the switching probability $P_{SW}=\frac{M_{SW}}{M}$ corresponding to the applied $I_{bias}$. By sweeping the $I_{bias}$ amplitude and repeating the above sequence, we measure a complete switching histogram, $P_{SW}$ \textit{vs} $I_{bias}$. The $P_{SW}=50\%$ bias current is called the switching current of the circuit, $I_{SW}$.

The choice of the read-out junction critical current $I^{RO}_{C}$ for an optimal measurement of $i_{S}$ is not straightforward. On the one hand one would like to have $I^{RO}_{C}\gg i_{S}$, but on the other hand the width of the switching histograms $w$ increases with $I^{RO}_{C}$ and so does the probabilistic noise in the measurement $w/\sqrt{M}$. For reasonable measuring time scales, the number of current steps $M$ is limited to values of about $10^{4}$. If we want to measure supercurrents for the SQUID chain in the range of $1nA$, $I^{RO}_{C}$ needs to be in the range of $100nA$. We have used a read-out junction with a critical current $I^{RO}_{C}=330nA$ which offers a good trade off.

\vspace{0.3cm}
\textbf{Acknowledgements}
\vspace{0.3cm}

We thank B. Dou\c{c}ot, D. Est\`{e}ve, F. Hekking and L. Ioffe for fruitful discussions. We are grateful to the team of the Nanofab facility in Grenoble for their technical support in the sample fabrication. Our research is supported by the European STREP MIDAS and the French ANR "QUANTJO".

\end{document}